\documentclass[12pt,a4paper]{article}

\usepackage[usenames]{color}
\usepackage{hyperref}
\usepackage{epsfig,amsmath,wrapfig,amsthm,dsfont}
\usepackage{mathrsfs}
\usepackage{amsfonts}
\usepackage{authblk}

\usepackage{graphicx}
\usepackage{subcaption}
\usepackage{natbib}
\usepackage{bm}

\usepackage{booktabs}

\usepackage{graphicx}

\usepackage[paperwidth=8.5in, paperheight=11in, margin=1in]{geometry}
\usepackage{setspace}
\def\boldfacefake#1{\kern-4pt
   \hbox{ \mathsurround=0pt
   \hbox to 0.4pt{$#1$\hss}\hbox to 0.4pt{$#1$\hss}\hbox {$#1$}}}

\newcommand{\btable}{\begin{table}[h]\centering}
\newcommand{\etable}{\end{table}}
\newcommand{\bt}{\begin{parag}\small \let\b=\nsb \let\sb=\nssb \begin{tabular}}
\newcommand{\et}{\end{tabular}\let\b=\nb \let\sb=\nsb\end{parag}}

\newenvironment{parag}{\par}{\par}

\newcommand{\be}{\begin{eqnarray}}
\newcommand{\ee}{\end{eqnarray}}
\newcommand{\ba}{\begin{eqnarray*}}
\newcommand{\ea}{\end{eqnarray*}}

\newtheorem{theorem0}{Theorem}
\newtheorem{lemma0}{Lemma}
\newtheorem{remark0}{Remark}
\newtheorem{fact0}{Fact}
\newtheorem{example0}{Example}
\newtheorem{definition0}{Definition}
\newtheorem{corollary0}{Corollary}
\newtheorem{proposition0}{Proposition}
\newtheorem{algorithmY}{Algorithm}

\newcommand{\reals}{\mbox{\rm I\kern-.20em R}}
\newcommand{\sreals}{\mbox{\small \rm I\kern-.20em R}}

\DeclareFontFamily{OT1}{pzc}{}
\DeclareFontShape{OT1}{pzc}{m}{it}{<-> s * [1.10] pzcmi7t}{}
\DeclareMathAlphabet{\mathpzc}{OT1}{pzc}{m}{it}

\makeatletter
\renewcommand*{\@fnsymbol}[1]{\ensuremath{\ifcase#1\or *\or *\or *\or
   \mathsection\or \mathparagraph\or \|\or **\or \dagger\dagger
   \or \ddagger\ddagger \else\@ctrerr\fi}}
\makeatother

\makeatletter
\if@compatibility
\renewenvironment{titlepage}
    {%
      \if@twocolumn
        \@restonecoltrue\onecolumn
      \else
        \@restonecolfalse\newpage
      \fi
      \thispagestyle{empty}%
    }%
    {\if@restonecol\twocolumn \else \newpage \fi
    }
\else
\renewenvironment{titlepage}
    {%
      \if@twocolumn
        \@restonecoltrue\onecolumn
      \else
        \@restonecolfalse\newpage
      \fi
      \thispagestyle{empty}%
      \setcounter{page}\@ne
    }%
    {\if@restonecol\twocolumn \else \newpage \fi
     \if@twoside\else
     \fi
    }
\fi
\makeatother

\title{
Association Analysis of Common and Rare SNVs using Adaptive Fisher Method to Detect Dense and Sparse Signals 
}

\author[1]{Xiaoyu Cai}
\author[1]{Lo-Bin Chang}
\author[2]{Chi Song\thanks{Correspondence to: Chi Song, College of Public Health, Division of Biostatistics, The Ohio State University, 1841 Neil Ave., 208E Cunz Hall, Columbus, OH 43210. E-mail: song.1188@osu.edu}} 
\affil[1]{Department of Statistics, The Ohio State University} 
\affil[2]{College of Public Health, Division of Biostatistics, The Ohio State University}

\begin{document}
\begin{titlepage}
\maketitle
\textbf{Running Title: Association Analysis of SNVs using AF}
\end{titlepage}

\begin{abstract}
The development of next generation sequencing (NGS) technology and genotype imputation methods enabled researchers to measure both common and rare variants in genome-wide association studies (GWAS).  Statistical methods have been proposed to test a set of genomic variants together to detect if any of them is associated with the phenotype or disease.  In practice, within the set of variants, there is an unknown proportion of variants truly causal or associated with the disease.  Because most developed methods are sensitive to either the dense scenario, where a large proportion of the variants are associated, or the sparse scenario, where only a small proportion of the variants are associated, there is a demand of statistical methods with high power in both scenarios.  In this paper, we propose a new association test (weighted Adaptive Fisher, wAF) that can adapt to both the dense and sparse scenario by adding weights to the Adaptive Fisher (AF) method we developed before.  Using both simulation and the Genetic Analysis Workshop 16 (GAW16) data, we have shown that the new method enjoys comparable or better power to popular methods such as sequence kernel association test (SKAT and SKAT-O) and adaptive SPU (aSPU) test.
\end{abstract}


\newpage



\section*{Introduction}
\label{sec:introduction}
Single nucleotide variants (SNVs) are a type of chromosome variants where the DNA sequence of an individual is different from the reference genome  on only one nucleotide.  Before the era of next generation sequencing (NGS), SNP array technology was used to obtain the genotypes of common SNVs with minor allele frequencies (MAFs) larger than certain cutoff (e.g. 1\% or 5\%, a.k.a single nucleotide polymorphisms or SNPs).   Over the past decades, genome-wide association studies (GWASs) have been successfully conducted to discover many disease-associated common SNVs with relatively large minor allele frequencies (MAFs) \citep{gwascatalog2013, gwascatalog2017}.  Despite the success of GWAS, the common SNVs detected through this procedure sometimes account for only a small proportion of the heritability, which is know as the problem of ``missing heritability'' \citep{manolio2009finding}.  This problem promotes the researchers to seek heritability outside of the controversial common disease-common variant hypothesis, which is the fundamental of GWAS based on common SNVs, but to seek ``missing heritability'' in rare SNVs \cite{schork2009common}.  Rare SNVs (a.k.a rare variants) are SNVs with low MAFs (often $<1\%$ or $<5\%$).  Comparing to common SNVs, the number of rare SNVs is much larger, and their locations on the human genome is often unknown before genotyping all the study samples, which makes DNA hybridization-based genotyping technology (e.g. SNP array) inapplicable in genotyping of rare SNVs.  Thanks to the advent of NGS, researchers now are enabled to reliably measure rare SNVs.  Furthermore, because of the development of fast imputation tools \citep{das2016next} and the 1000 Genomes project \citep{10002015global}, rare SNVs can be imputed for old GWASs where only common SNVs were measured.  This helps recycle and add value to the numerous GWASs that are conducted for many complex human diseases and are available on public domain.

However, the technology advancement in genotyping of rare SNVs also presents several statistical challenges for the association analysis method development.  First, because of the small MAFs of the rare variants, the statistical power of traditional association methods are very low when applied to detect association between rare variants and the disease outcome.  Second, because the number of SNVs including both common and rare variants are significantly larger than the number of common variants (often more than 100 times larger), the multiple comparison issue is more severe \citep{song2014tarv}.  Therefore, it would be powerless if association analysis were performed on each single SNV separately. A commonly used solution to these issues was to perform the association analysis on SNV sets, where multiple SNVs grouped together based on their locations on the genome.  SNVs on or close to a gene are often grouped together into one SNV set.  However, the traditional statistical testing methods such as score test or likelihood ratio test used in multivariate generalized linear model (GLM) are not powerful enough when many variants are included in the SNV set. As shown by \citet{fan1996test}, the tests based on $\chi^2$ distribution will have no power when the signal is weak or rare as the degree of freedom increases.  To solve this problem, three categories of approaches have been proposed, all of them essentially reduced the degree of freedom in some way to boost the statistical power.

The first category is burden tests, which collapse rare variants into genetic burdens, then test the effects of the genetic burden.  CAST \citep{morgenthaler2007strategy}, CMC \citep{li2008methods} and wSum \citep{madsen2009groupwise} all belong to this category.  By combining multiple rare variants into a single measurement of genetic burden, these methods essentially reduced the number of parameters to test down to one, which is equivalent to reducing the degree of freedom of the $\chi^2$ test statistic to one.  Despite the popularity of this type of methods, the traditional way of calculating genetic burden often ignores the fact that different variants may have opposite effects on the same outcome.  Simply pooling or summing the variants together may cause the opposite effects to cancel out, therefore reduce the statistical power.  A solution is to calculate genetic burden adaptively based on evidence provided by the data.  For example, \citet{price2010pooled} proposed to adjust minor allele frequency (MAF) threshold for the pooling step based on data.  \citet{han2010data} and \citet{hoffmann2010comprehensive} proposed to adaptively choose the sign and magnitude of the weight in the collapsing step to calculate genetic burdens.  TARV \citep{song2014tarv} can also be viewed as this type of method because it adaptively combines multiple variants into a ``super variant'' based on the strength of evidence provided by each single variant.

The second category of methods is quadratic tests which often base on test of variance component in mixed effect models.  The well-known SKAT \citep{wu2011rare} belongs to this category.  By assuming the effect of each variant being random, SKAT tests whether the variance of the random effects is zero.  The test statistic can be approximated by a $\chi^2$ distribution with a degree of freedom much smaller than the degree of freedom in the likelihood ratio test (or Rao's score test) in the fixed effect models.  SKAT can also test non-linear effects by adopting an arbitrary kernel matrix.  SKAT was also extended to accommodate multiple candidate kernels \citep{wu2013kernel}, to jointly test rare and common variants \citep{ionita2013sequence}, and to apply on family data \citep{chen2013sequence}.  Some other popular methods, such as C-alpha \citep{neale2011testing} and SSU \citep{pan2009asymptotic} can be viewed as special cases of SKAT.

The third category is functional analysis.  Because the genomic variants within the same gene are often highly correlated due to linkage disequibrillium (LD), this category of methods treat them as discrete realizations of a hidden continuous function on the genome.  Both the variants and their coefficients can then be decomposed in the functional space.  Since the number of functional bases used is generally smaller than the number of variants, this is equivalent to a dimensional reduction method which also reduces the degree of freedom of the association test.  Different methods under this category has been proposed utilizing different basis including functional principle component basis \citep{luo2011association}, B-spline basis \citep{luo2012quantitative,fan2013functional}, and Fourier basis \citep{fan2013functional}.

In addition to these three categories of methods, effort has also been made to combine multiple testing methods into one single test.  For example, the popular SKAT-O \citep{lee2012optimal} is a combination of variance component test (SKAT) and burden test.  Similarly, \citet{derkach2013robust} proposed to combine variance component test and burden test using Fisher's method or minimal P-value.

It should be noted that the power of aforementioned methods relies on the proportion of variants which truly associate with the disease outcome.  Under the alternative hypothesis -- when the null hypothesis of no association is untrue, all three types of methods assume that every SNVs included in the test has some nonzero effect more or less.  Specifically, burden tests assume the effects of the variants are proportional to each other, with the proportion predefined by the weights used to calculate the genetic burden; variance component tests assume the random effects of the combined variants share a common variance component, which if is not zero implies all the random effects are nonzero; and the functional analysis based methods, tests whether any functional basis (a weighted sum of variants) has a nonzero effect, which in turn implies nonzero effects for all or most of the variants.  The type I error of these methods is not affected by violation of this assumption of alternative hypothesis, which does not undermine their validity.  However, under alternative hypothesis where not all of the effects are nonzero, especially when only a small proportion of variants have nonzero effects, the statistical power of these tests will be suboptimal.  Therefore there is a demand for statistical methods that can adapt to the proportion of variants with nonzero effects.  For the ease of discussion, we call the scenario where this proportion is large as the dense scenario, and call the scenario where this proportional is small as the sparse scenario.  For this purpose, \citet{pan2014powerful} proposed an adaptive test named aSPU which has strong statistical power in both the dense and sparse scenarios.  This aSPU can also be viewed as a combination of SKAT (with linear kernel) and other tests including burden test.  \citet{barnett2014analytical} suggested that Higher Criticism (HC) can be another potential powerful test that can adaptively detect both dense and sparse signals.  Previously, we proposed Adaptive Fisher (AF) method \cite{song:min:zhang:aoas} and illustrated in simulation that AF is a very powerful method to detect the mixture distribution in both dense and sparse scenarios, and it can be much more powerful than HC with finite sample.  Therefore, we propose to use AF to detect disease associated SNV sets, and compare to existing methods in the following section.

\section*{Methods}

Suppose a trait for $n$ independent subjects $\boldsymbol{Y}=(Y_{i1},...,Y_{in})^T$ are observed. $\boldsymbol{G}_i=(G_{i1},...,G_{iK})^T$ denotes the genotypes of $K$ SNVs in a chromosomal region (e.g. a gene) for subject $i$, where $G_{ik}=0,1,2$ represents the number of minor alleles at locus $k$ of subject $i$. We model associations between the trait and SNVs with the following generalized linear model
\begin{equation} \label{eq:1}
h\Big(E(Y_i)\Big)=\beta_0+\sum_{k=1}^K \beta_k G_{ik},
\end{equation}
where $\boldsymbol{\beta}=(\beta_1,...,\beta_K)^T$ is the vector of SNV effects, and vector $\boldsymbol{\alpha}=(\alpha_1,...,\alpha_J)^T$ contains covariate effects. $h(\cdot)$ is taken as the logit link function for binary traits (e.g. diseased or nondiseased) or the identity link function for continuous traits (e.g. blood pressure, height, etc.). If $J$ covariates $\boldsymbol{C}_i=(C_{i1},..., C_{iJ})^T$, $i=1,2,...,n$ are also observed for each subject, the model can be extended as
\begin{equation} \label{eq:2}
h\Big(E(Y_i)\Big)=\beta_0+\sum_{k=1}^K \beta_k G_{ik}+\sum_{j=1}^J\alpha_j C_{ij}
\end{equation}
Determining whether there is an association between the trait and any SNV is equivalent to testing the following hypotheses,
\begin{equation} \label{eq:3}
H_0: \boldsymbol{\beta}=\boldsymbol{0} \quad \text{versus} \quad H_1: \boldsymbol{\beta} \neq \boldsymbol{0}. 
\end{equation}

The proposed adaptive fisher tests involve the score statistics $\boldsymbol{U}=(U_1,...,U_K)^T$. For model \eqref{eq:1},
\begin{equation} \label{eq:4}
\boldsymbol{U}=\sum_{i=1}^n (Y_i-\bar{Y})\boldsymbol{G}_i
\end{equation}
and its estimated covariance matrix under $H_0$ is given by
\begin{equation} \label{eq:5}
\boldsymbol{V}=\widehat{Cov}(U|H_0)=\bar{Y}(1-\bar{Y})\sum_{i=1}^n (\boldsymbol{G}_i-\bar{\boldsymbol{G}})(\boldsymbol{G}_i-\bar{\boldsymbol{G}})^T.
\end{equation}
for the binary traits, and
\begin{equation} \label{eq:6}
\boldsymbol{V}=\widehat{Cov}(U|H_0)=\hat{\sigma}_1^2 \sum_{i=1}^n (\boldsymbol{G}_i-\bar{\boldsymbol{G}})(\boldsymbol{G}_i-\bar{\boldsymbol{G}})^T
\end{equation}
for the continuous traits, 
where $\bar{Y}=\frac{1}{n} \sum_{i=1}^n Y_i$, $\hat{\sigma}_1^2=\frac{1}{n-1} \sum_{i=1}^n (Y_i-\bar{Y})^2$ and $\bar{\boldsymbol{G}}=(\bar{G}_{\cdot 1}, ..., \bar{G}_{\cdot K})^T$ with $\bar{G}_{\cdot k}=\frac{1}{n} \sum_{i=1}^n G_{ik}$. 
For model \eqref{eq:2},
\begin{equation} \label{eq:7}
\boldsymbol{U}=\sum_{i=1}^n (Y_i-\hat{\mu}_{Y_i})(\boldsymbol{G}_i-\hat{\boldsymbol{G}_i}),
\end{equation}
for the binary traits, 
\begin{equation} \label{eq:8}
\boldsymbol{V}=\widehat{Cov}(U|H_0)=\hat{\sigma}_2^2 \sum_{i=1}^n (\boldsymbol{G}_i-\hat{\boldsymbol{G}_i})(\boldsymbol{G}_i-\hat{\boldsymbol{G}_i})^T,
\end{equation}
and for the continuous traits,
\begin{equation} \label{eq:9}
\boldsymbol{V}=\widehat{Cov}(U|H_0)=\hat{\sigma}_3^2 \sum_{i=1}^n (\boldsymbol{G}_i-\hat{\boldsymbol{G}_i})(\boldsymbol{G}_i-\hat{\boldsymbol{G}_i})^T,
\end{equation}
where $\hat{\mu}_{Y_i}=h^{-1}(\hat{\beta}_0+\sum_{j=1}^J\hat{\alpha}_j C_{ij})$ with $\hat{\beta}_0$ and $\hat{\alpha}_j, \ j=1,2,...,J$ being the maximum likelihood estimators, $\hat{\boldsymbol{G}_i}=(\hat{G}_{i1},...,\hat{G}_{ik})$ with $\hat{G}_{ik}$ being the predictive value of $G_{ik}$ from a linear regression model with covariates as predictors,
$\hat{\sigma}_2^2=\frac{1}{n} \sum_{i=1}^n (\hat{\mu}_{Y_i}(1-\hat{\mu}_{Y_i})$, and $\hat{\sigma}_3^2=\frac{1}{n-1} \sum_{i=1}^n (Y_i-\hat{\mu}_{Y_i})^2$.

\subsection*{Adaptive Fisher Method}
Let the standardized score statistics be $\tilde{U}_k=U_k/\sqrt{V_{kk}}$, where $V_{kk}$ is the $k^{\text{th}}$ diagonal element of $\boldsymbol{V}$. If $\beta_k$ is tested marginally, the P-value for this marginal score test is $p_k=2\big(1-\Phi(|\tilde{U_k}|)\big)$, $k=1,2,...,K$, as $\tilde{U}_k$ is asymptotically $N(0,1)$ distributed under $H_0$. Let 
\begin{equation} \label{eq:10}
R_k=-\log \ p_k.
\end{equation}
Order $R$'s in descending order $R_{(1)} \geq \cdots \geq R_{(K)}$.
Let $\boldsymbol{S}=(S_1,...,S_K)^T$ be the partial sums of $R_{(1)}, ..., R_{(K)}$,
\begin{equation} \label{eq:11}
S_k=\sum_{l=1}^k R_{(l)}.
\end{equation}
For each $S_k$, $k=1,2,...,K$, we calculated its P-value by
\begin{equation} \label{eq:12}
P_{s_k}=\text{Pr}(S_k \geq s_k),
\end{equation}
where $s_k$ is be observed value of $S_k$.
The AF test is based on the AF statistic below
\begin{equation} \label{eq:13}
T_{\text{AF}}=\min_{1 \leq k \leq K} P_{s_k}.
\end{equation}

\subsection*{Weighted Adaptive Fisher Method}
SNVs can be weighed differently when taking the partial sums. Suppose $\boldsymbol{w}=(w_1,...,w_K)$ are weights of the $K$ SNVs in a genetic region. Define
\begin{equation} \label{eq:14}
X_k=w_k R_k.
\end{equation}
Order $X_1,...,X_K$ in descending order $X_{(1)},...,X_{(K)}$. Let $\boldsymbol{S^*}=(S^*_1,...,S^*_K)^T$ be the partial sums of $X_{(1)}, ..., X_{(K)}$
\begin{equation} \label{eq:15}
S^*_k=\sum_{l=1}^k X_{(l)}.
\end{equation}
Similar to \eqref{eq:12}, the P-value of $s^*_k$ (observed value of $S^*_k$), $P_{s^*_k}= \text{Pr}(S^*_k \geq s^*_k)$, and the weighted AF (wAF) statistic is defined by
\begin{equation} \label{eq:16}
T_{\text{wAF}}=\min_{1 \leq k \leq K} P_{s^*_k}.
\end{equation}

\subsection*{Computation}

The following permutation procedure is needed for accessing $P_{s_k}$ ($P_{s^*_k}$) in \eqref{eq:12} and finding the null distributions of $T_{\text{AF}}$ in \eqref{eq:13} and $T_{\text{wAF}}$ in \eqref{eq:16}. Here the weighted version for model \eqref{eq:1} is used as an example. The unweighted method can be treated as a special case of all weights being equal.

\begin{itemize}
	\item[1.] Calculate residuals by $e_i=Y_i-\bar{Y}$, $i=1,2,...,n$.
	\item[2.] Permute $e_i$'s for a large number $B$ times to obtain $\boldsymbol{e}^{(b)}=(e^{(b)}_1,...,e^{(b)}_n)^T$, $b=1,2,...,B$ where $(e^{(b)}_1,...,e^{(b)}_n)^T$ is a permutation of $\boldsymbol{e}^{(0)}=(e_1,...,e_n)^T$.
	\item[3.] For each $\boldsymbol{e}^{(b)}$, calculate $\boldsymbol{U}^{(b)}=(U^{(b)}_1,...,U^{(b)}_K)^T=\sum_{i=1}^n e_i \boldsymbol{G}_i$ and $\boldsymbol{p}^{(b)}=(p^{(b)}_1,...,p^{(b)}_K)^T$ with $p^{(b)}_k=2\big(1-\Phi(|U^{(b)}_k/\sqrt{V_{kk}}|))$. Then follow equations \eqref{eq:10}, \eqref{eq:14} and \eqref{eq:15} to get $\boldsymbol{S}^{*(b)}=(S^{*(b)}_1,...,S^{*(b)}_K)^T$, $b=0,1,2,..,B$.
	\item[4.] For a fixed $b^* \in \{0,1,2,...B\}$,
	$$P^{(b^*)}_{S^*_k}=\frac{1}{B+1} \sum_{b=0}^B \mathbb{I} \{S^{*(b)}_k \geq S^{*(b^*)}_k\}.$$
	\item[5.] For each $\boldsymbol{S}^{*(b)}$, $T_{\text{wAF}}^{(b)}=\min_{1 \leq k \leq K} P_{S^*_k}^{(b)}$, $b=0,1,2,...,B$.
	\item[6.] The P-value of wAF test can be approximated by
	$$\widehat{\text{Pr}}\{ T_{\text{wAF}} \leq T_{\text{wAF}}^{(0)}|H_0\}=\frac{1}{B+1} \sum_{b=1}^B \mathbb{I}\{ T_{\text{wAF}}^{(b)} \leq T_{\text{wAF}}^{(0)}\}$$
	where $T_{\text{wAF}}^{(0)}=\min_{1 \leq k \leq K} P_{S^*_k}^{(0)}$ is the observed value of the wAF statistic and $\mathbb{I}(\cdot)$ is the indicator function.
\end{itemize}
The above procedure applies to model \eqref{eq:2} as well, with $e_i$ being replaced by $e_i=Y_i-\hat{\mu}_{Y_i}$, $i=1,2,...,n$ as the only change. 

\section*{Results}
We evaluate our wAF method using both simulation and the Genetic Analysis Workshop 17 (GAW17) data.  The performance of our method is compared to SKAT, SKAT-O, aSPU and Min-P (which takes the minimal P-value of all the combined variants as the test statistic).

\subsection*{Simulation Studies}

Simulation studies are conducted to evaluate the performance of wAF test under both dense and sparse scenarios. 
The simulation analysis was performed following the simulation framework of \citet{pan2014powerful}. 





 First, genotypes $\boldsymbol{G}_i=(G_{i1},...,G_{iK})^T$, $i=1,2,...,n$ are simulated by the following steps, similar to the simulation setups in \citet{pan2014powerful}.

\begin{itemize}
	\item[1.] Generate $\boldsymbol{Z_1}=(Z_{11},...,Z_{1K})^T$ and $\boldsymbol{Z_2}=(Z_{21},...,Z_{2K})^T$ independently from a multivariate normal distribution $N(\boldsymbol{0},\boldsymbol{A})$. $\boldsymbol{A}$ has a first-order autoregressive (AR$(1)$) covariance structure with the $(k,k')^{\text{th}}$ element $A_{kk'}=c^{|k-k'|}$. $c$ is chosen to be $0.9$ to give close loci a higher correlation and distant loci a lower correlation.
	\item[2.]  Randomly sample the minor allele frequencies (MAF's) by first generating $\log(\text{MAF})$'s from $U(\log(0.001),\log(0.05))$ and then exponentiating them back to MAFs.\footnote{Because of the logarithm, this MAF sampling algorithm often samples small MAF's and therefore yields more rare variants.}  Set $G_{ik}=\mathbb{I}(\Phi(Z_{1k}) \leq \text{MAF}_k)+\mathbb{I}(\Phi(Z_{2k}) \leq \text{MAF}_k)$, $k=1,...,K$.
	\item[3.] Repeat step 1 and 2 $n$ times to generate genotypes for all subjects. 
\end{itemize}

For genotype effects $\boldsymbol{\beta}=(\beta_1,...,\beta_K)^T$, randomly sample $\pi K$ effects to be nonzero, whose values are sampled from a uniform distribution within $[-\delta,\delta]$, while keep the other $(1-\pi) K$ effects remain zeros. Trait of $n=1,000$ subjects are generated from model \eqref{eq:1}.

The weights of the wAF test are chosen to be $w_k=\sqrt{MAF_k(1-MAF_k)}$, $k=1,2,..,K$. The weights of SKAT and SKAT-O are chosen to be flat with $w_k=1$, $k=1,2,...,K$, so that SKAT is equivalent to SSU \cite{wu2011rare}. The significance level is set to be 0.05 for every test. All simulation results are based on $5,000$ replicates. 


\subsubsection*{Binary Traits}

When generating binary trait, $h(\cdot)$ is taken to be the logit link function. We increase the number of SNVs, $K$, from $50$ to $500$ with an increment $50$, while hold the effect proportion $\pi$ and  the effect size $\delta$ constant. For the dense scenario, $\pi=20 \%$ and $\delta=0.25$. For the sparse scenario, $\pi=2 \%$ and $\delta=1$. Figure 1 shows that the wAF test results in large powers for the both dense and sparse scenarios.  Specifically, in the dense scenario, wAF and SKAT have the highest power.  SKAT-O and aSPU are slightly less powerful than SKAT and wAF.  Min-P, on the other hand, is much less powerful than the other methods.  For the sparse scenario, Min-P is the most powerful method.  Our wAF has the second highest power which is about 5\% to 10\% higher than the other methods including SKAT, SKAT-O and aSPU.  For all these compared methods, the type I errors are well-controlled empirically as shown in the supplementary table 1.


\begin{figure}[]
	\centering
	\begin{subfigure}[h]{0.49\textwidth}   
		\centering 
		\includegraphics[width=\textwidth]{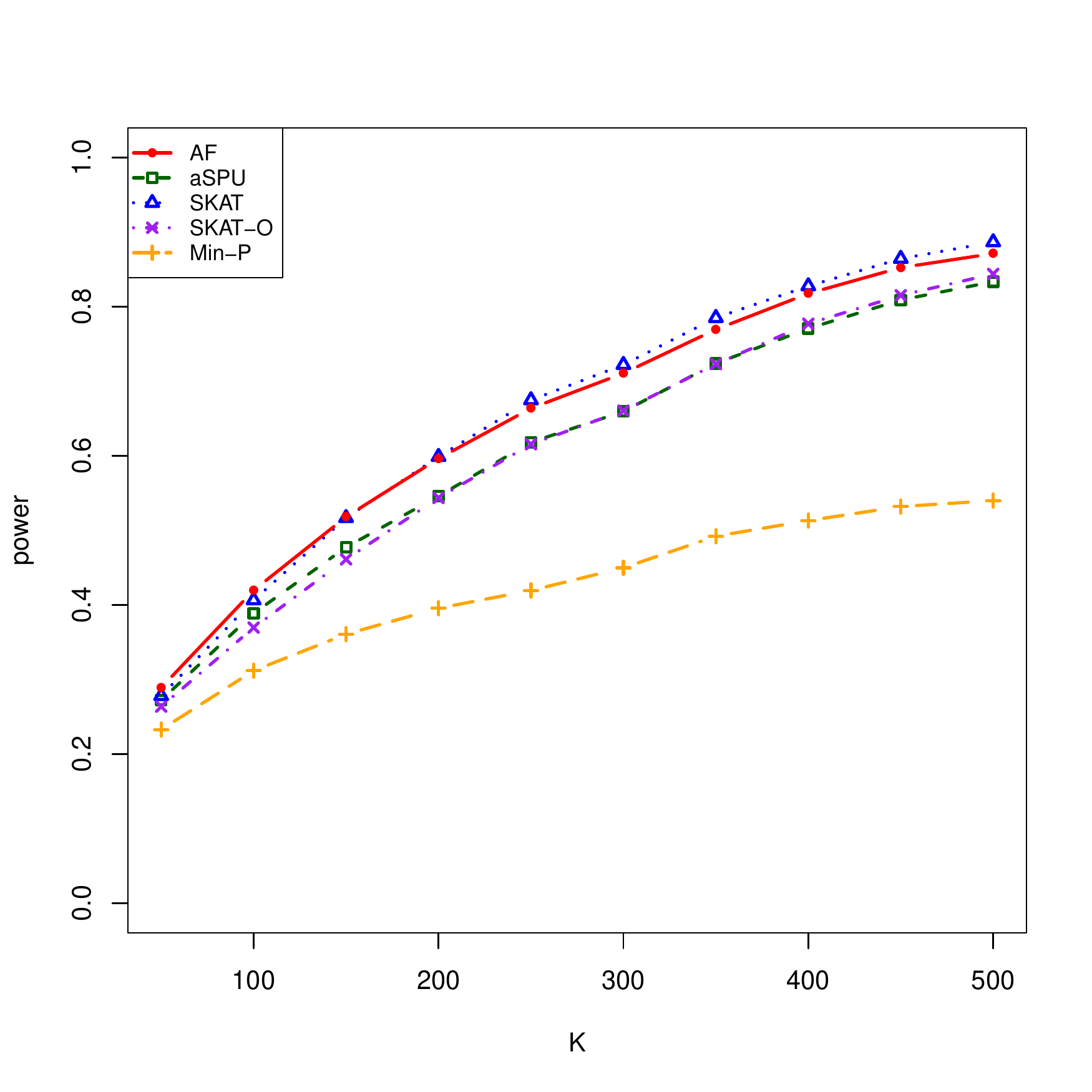}
		\caption[]%
		{{\small Dense Scenario}}    
	\end{subfigure}
	\begin{subfigure}[h]{0.49\textwidth}
		\centering
		\includegraphics[width=\textwidth]{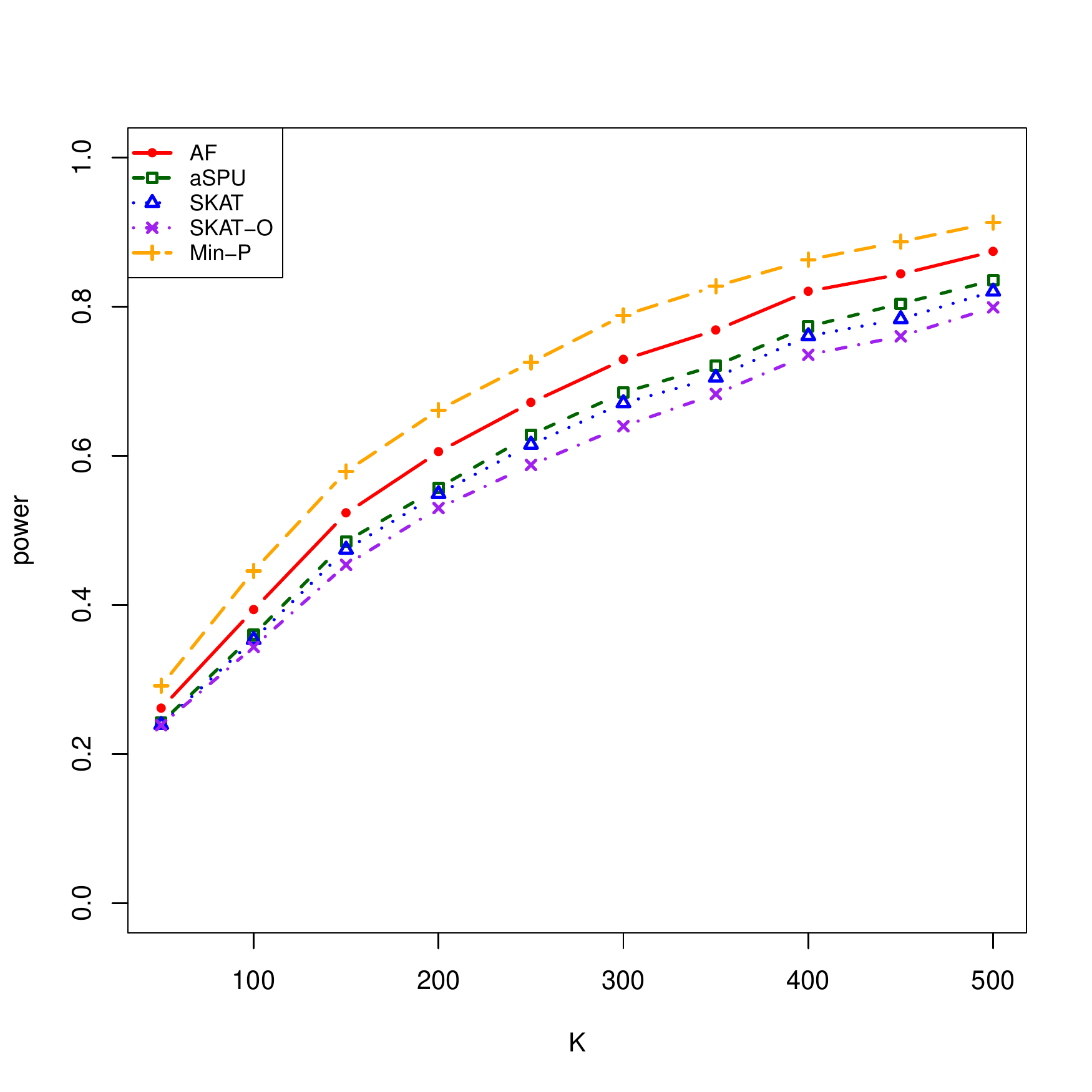}
		\caption[]%
		{{\small Sparse Scenario}}    
	\end{subfigure}
	
	\caption[]
	{\small Comparison of empirical powers of five methods for binary trait. (a) Power against varying number of loci $K$ in the dense scenario with effect proportion $\pi=20 \%$ and effect size $\delta=0.25$. $K$ varies from $50$ to $500$. 
	(b) Power against varying number of loci $K$ in the sparse scenario with effect proportion $\pi=2 \%$ and effect size $\delta=1$. $K$ varies from $50$ to $500$. 
	} 
\end{figure}


\subsubsection*{Continuous Traits}

When generating continuous trait, $h(\cdot)$ is taken to be the identity link function and random errors are standard normal random variables. Again, $K$ is increased from $50$ to $500$ with an increment $50$, while $\pi$ and  $\delta$ are held constants. For the dense scenario, $\pi=20 \%$ and $\delta=0.15$. For the sparse scenario, $\pi=2 \%$ and $\delta=0.5$.  Based on power curves in Figure 2, the wAF test performs relatively well for the both dense and sparse scenarios similar to what we have seen in the binary trait.  In dense scenario, wAF and SKAT enjoys the highest power, which is slightly better than aSPU and SKAT-O, and much better than Min-P.  Whereas in the sparse scenario, Min-P is the most powerful method, seconded by wAF, and wAF has higher power than aSPU, SKAT and SKAT-O.  Similar to the binary traits, all type I errors are well-controlled empirically as seen in the supplementary table 1.


\begin{figure}[]
	\centering
	\begin{subfigure}[h]{0.49\textwidth}   
		\centering 
		\includegraphics[width=\textwidth]{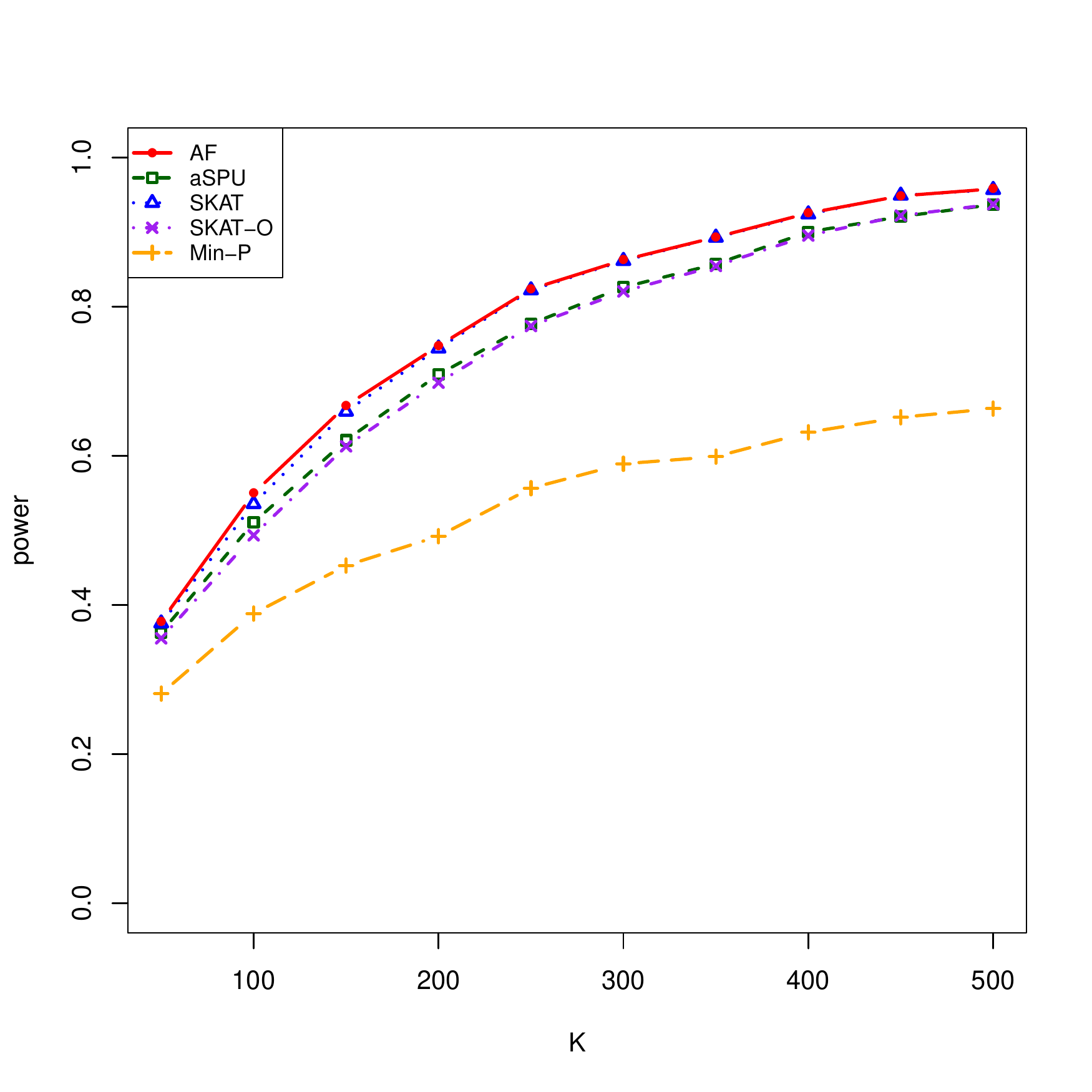}
		\caption[]%
		{{\small Dense Scenario}}    
	\end{subfigure}
	\begin{subfigure}[h]{0.49\textwidth}
		\centering
		\includegraphics[width=\textwidth]{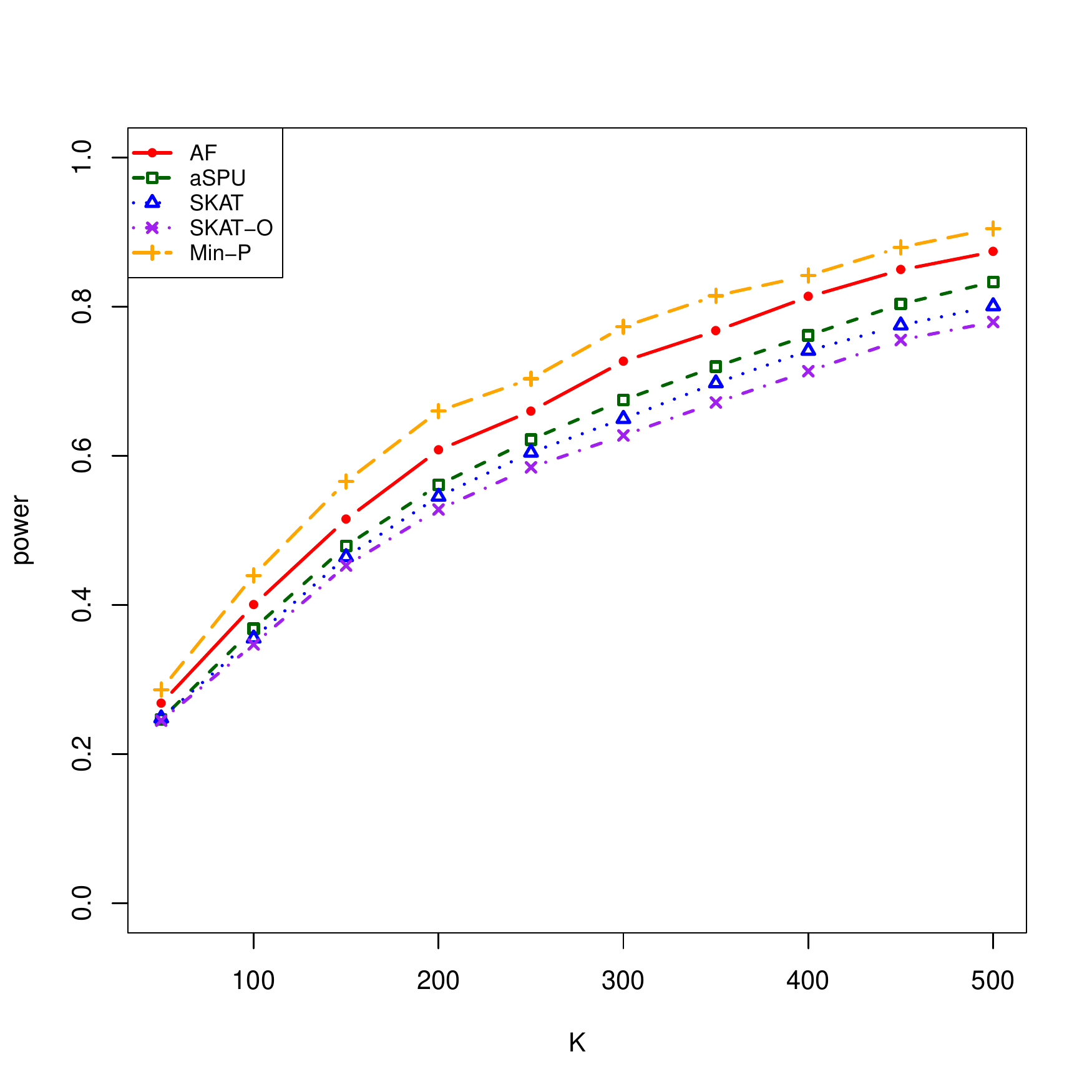}
		\caption[]%
		{{\small Sparse Scenario}}    
	\end{subfigure}
	
	\caption[]
	{\small Comparison of empirical powers of five methods for continuous trait. (a) Power against varying number of loci $K$ in the dense scenario with effect proportion $\pi=20 \%$ and effect size $\delta=0.15$. $K$ varies from $50$ to $500$. 
	(b) Power against varying number of loci $K$ in the sparse scenario with effect proportion $\pi=2 \%$ and effect size $\delta=0.5$. $K$ varies from $50$ to $500$. 
	} 
\end{figure}


\subsection*{Application to GAW17 data}

The above five methods (wAF, aSPU, SKAT, SKAT-O and Min-P) are applied to Genetic Analysis Workshop 17 (GAW17) mini-exome simulation data in Almasy \textit{et al.} (2011) \cite{almasy2011genetic}. $24,487$ SNVs in $3,205$ genes from $697$ subjects are genotyped. If only rare variants (with MAFs no larger than $1 \%$) are considered, there are $18,131$ rare variants in $2,476$ genes (that contain at least $1$ rare variant). $200$ sets of binary traits are simulated based on genotypes and three covariates: age, gender and smoking status.  When only rare variants are considered, $35$ causal genes have effects on the trait.

We follow the same procedure as \citet{pan2014powerful}. We apply the five tests on each of the causal genes separately with gene-wise significance level $0.05$, and estimate the corresponding powers, using the 200 sets of phenotypes.
Removing genes for which all tests have power lower than $10 \%$, the estimated powers for the remaining $17$ genes are shown in Table \ref{tb3}. 

\begin{table}[]
	\centering
	\caption{Estimated Power for Some Causal Genes}
	\label{tb3}
	\begin{tabular}{@{}lccccc@{}}
		\toprule
		\multicolumn{1}{c}{\textbf{Gene}} & \textbf{wAF} & \textbf{aSPU} & \textbf{SKAT} & \textbf{SKATO} & \textbf{Min-P} \\ \midrule
		PIK3C2B                           & 0.440       & 0.625         & 0.430         & 0.575          & 0.270          \\
		BCHE                              & 0.245       & 0.210         & 0.155         & 0.145          & 0.165          \\
		KDR                               & 0.360       & 0.295         & 0.385         & 0.400          & 0.050          \\
		VNN1                              & 0.275       & 0.255         & 0.195         & 0.180          & 0.155          \\
		INSIG1                            & 0.015       & 0.015         & 0.225         & 0.225          & 0.000          \\
		LPL                               & 0.135       & 0.125         & 0.120         & 0.105          & 0.055          \\
		PTK2B                             & 0.065       & 0.070         & 0.065         & 0.085          & 0.110          \\
		PLAT                              & 0.145       & 0.145         & 0.120         & 0.180          & 0.060          \\
		VLDLR                             & 0.110       & 0.085         & 0.100         & 0.085          & 0.090          \\
		SIRT1                             & 0.110       & 0.090         & 0.095         & 0.090          & 0.035          \\
		VWF                               & 0.060       & 0.020         & 0.015         & 0.030          & 0.135          \\
		FLT1                              & 0.155       & 0.115         & 0.140         & 0.160          & 0.045          \\
		SOS2                              & 0.270       & 0.220         & 0.270         & 0.210          & 0.100          \\
		HSP90AA1                          & 0.345       & 0.145         & 0.255         & 0.190          & 0.290          \\
		SREBF1                            & 0.105       & 0.075         & 0.075         & 0.070          & 0.105          \\
		PRKCA                             & 0.030       & 0.030         & 0.185         & 0.185          & 0.000          \\
		RRAS                              & 0.150       & 0.185         & 0.135         & 0.240          & 0.070          \\ \bottomrule
	\end{tabular}
\end{table}

\section*{Discussion}

Association analysis of SNV sets becomes the standard analysis approach in GWAS when rare variants are genotyped or imputed in the dataset.  However, when many SNVs are combined together into one omnibus test, the power of the statistical test often depends on the proportion of variants with nonzero effects and how these variants are combined.  Most current methods except aSPU is not adaptive to this proportion and only applies to either the dense or sparse scenario.  In this paper, we proposed a new adaptive method wAF as an alternative to aSPU with better or comparable power.  Based on simulation, we can see that for both dense and sparse scenarios, wAF outperformed aSPU in terms of statistical power for either binary or continuous outcome.  In the analysis of GAW17 data, we found that wAF sometimes outperformed and sometimes underperformed comparing to aSPU.  This is because in GAW17, the datasets were simulated such that all of the rare variants are risk factors, which means that the minor alleles always increase the risk of disease.  By having the SNVs with effects of the same direction (increase risk), burden tests become more favorable than variance component tests.  As shown by \citet{pan2014powerful}, SPU tests with odds powers take into account the direction of the effect, which can be considered as various types of burden tests.  Therefore, aSPU may enjoy the high power of burden tests, hence performed better than wAF in some genes of GAW17.  To improve the power for this situation when all or most of the causal variants have the same effect direction, we can use wAF to combine (1) the 2-sided P-values (as shown in this paper), (2) the 1-sided P-values on whether the variants are risk factors and (3) the 1-sided P-values on whether the variants are protective.  Then use the minimal P-value of (1), (2) and (3) as our test statistic.  However, because this alteration is very trivial and to avoid distracting the readers from our major methodology, we only illustrated (1) in this paper.

As stated in introduction, HC is another methods that can be used to combine marginal tests of each variant.  Although we did not explore the application of HC in SNV set analysis, \citet{barnett2017generalized} proposed a generalized higher criticism (GHC) based on HC.  They found that GHC was only powerful in sparse scenario but underperformed in dense scenario, and suggested that one may consider combining GHC and SKAT to boost power when we do not know which scenario the causal gene actually belongs to, which we believe is almost true for every real life problem.  This conclusion agrees with our previous findings about HC \citep{song:min:zhang:aoas}.

While comparing wAF and aSPU, we found that their test statistics can be written in the same general format.  For both methods, we can think the test statistic as adaptively chosen from a set of weighted sums with different weights.  The weighted sums in both methods can be written as $\sum_k v_c(\tilde{U}_k, G_k)w(G_k)f(\tilde{U}_k)$, where $v_c(\tilde{U}_k, G_k)$ is the $c$th adaptive weight function depends on the standardized score statistic and the genotype data for variant $k$, $w(G_k)$ is a non-adaptive weight only depends on the genotype data, and $f(\tilde{U}_k)$ is a transformation of the standardized score statistic.  We can show that for aSPU, $f(\tilde{U}_k)=\tilde{U}_k$, $w(G_k)=\mbox{sd}(G_k)$, and $v_c(\tilde{U}_k, G_k)=[w(G_k)f(\tilde{U}_k)]^{(c-1)}$ for $c\in \{1, 2, \ldots, 8, \infty\}$; for wAF, $f(\tilde{U}_k)=2[1-\Phi(|\tilde{U}_k|)]$, $w(G_k)=\sqrt{\mbox{MAF}_k(1-\mbox{MAF}_k)}\approx \mbox{sd}(G_k)/\sqrt{2}$, and $v_c(\tilde{U}_k, G_k)=I\{w(G_k)f(\tilde{U}_k)\ge [w(G)f(\tilde{U})]_{(c)}\}$ for $c\in \{1, 2, \ldots, K\}$, where $I\{\cdot\}$ is an indicator function $[\cdot]_{(c)}$ denotes the $c$th largest order statistics of the quantity inside the bracket.  By comparison, we can see that the major difference between aSPU and wAF is how we adaptively weight the test statistic: aSPU creates the weight by raising the statistics to different power, whereas wAF sequentially put a 0/1 weight based on the magnitude of the test statistics.  This comparison also reveals that although not explicitly mentioned, aSPU also weighs different variants based on their MAF using almost the same weight as we used in wAF.

Because permutation is needed for wAF, computational burden is a major weakness.  To improve computation speed, we adopt the same strategy as \citet{pan2014powerful} to run a hundred permutation first, then choose to increase number of permutation only for those with small P-values.  Theoretically, because sorting and order statistics are used in wAF, the computation complexity is higher than aSPU.  Specifically, because wAF need sorting and cumulative summation, our complexity is higher than aSPU by an order of $\log K$.  In practice, because $K$ is often fixed, the theoretical difference in computational complexity can be ignored.  In our future work, we plan to improve the computation of wAF by importance sampling.

In conclusion, we developed wAF, a powerful statistical method for SNV set association analysis that performs better than current available methods in both dense and sparse scenarios.

\section*{Acknowledgments}
The authors thanks Kellie Archer and Shili Lin for their helpful comments and \citet{OhioSupercomputerCenter1987} for the computational support.

\bibliographystyle{plainnat}


\bibliography{Ref}


\end{document}